\documentclass[
reprint,
superscriptaddress,
amsmath,amssymb,amsthm,
aps,
prl
]{revtex4-2}

\usepackage[utf8]{inputenc}
\usepackage[english]{babel}
\usepackage[T1]{fontenc}
\usepackage{tikz}
\usepackage{lipsum}
\usepackage{braket}
\usepackage{physics}
\usepackage{mathtools}
\usepackage{dsfont}
\usepackage{scalerel}
\usepackage{ulem}
\usepackage{cancel}
\usepackage{mathbbol}
\usepackage{ulem}

\usepackage{graphicx}
\usepackage{dcolumn}
\usepackage{bm}
\usepackage{hyperref}
\usepackage{footmisc}
\usepackage{hyperref}

\newcounter{detailfootnote}

\DeclareSymbolFontAlphabet{\amsmathbb}{AMSb}%

\newcommand{\beq}{\begin{equation} }
\newcommand{\eeq}{\end{equation} }
\newcommand{\bse}{\begin{subequations} }
\newcommand{\ese}{\end{subequations} }
\newcommand{\bea}{\begin{align} }
\newcommand{\eea}{\end{align} }
\newcommand{\tp}{\intercal}
\def\hid{\hat{\mathbb{1}}}

\def\calL{{\cal L}}

\def\calLp{{\cal L}^{\prime}}

\def\calH{{\cal H}}

\def\calHb{\bar{\cal H}}

\def\calLb{\bar{{\cal L}}}
\def\Lamb{\Lambda}

\def\ie{\textit{i.e., }}
\def\hA{\hat A}

\def\hB{\hat B}

\def\hT{\hat T}
\def\hTp{{\hat T}^\prime}

\def\hH{\hat H}
\def\hHb{\hat{ \bar{H}}}
\def\hHp{\hat H^{\prime}}

\def\hHb{\hat{H}_{\vecC}}
\def\hO{\hat O}
\def\hL{\hat L}
\def\hLp{\hat L^\prime}

\def\hsig{\hat{\sigma}}
\def\mW{\mathbb{W}}
\def\hrho{\hat \rho}

\def\alk{\alpha^{(k)}}

\def\alkb{{\alpha}^{*(k)}}
\def\mA{\mathbb{ A}}
\def\mAp{\mathbb{ A}^\prime}

\def\mD{\mathbb{D}}
\def\mDp{\mathbb{D}^\prime}
\def\mDt{\tilde{\mathbb{D}}}

\def\mO{\mathbb{O}}

\def\mf{\mathbb{f}}

\def\mC{\mathbb{C}}
\def\mCp{\mathbb{C}^\prime}
\def\mCt{\tilde{\mathbb{C}}}

\def\mGam{\mathbb{\Gamma}}
\def\mGamt{\tilde{\mathbb{\Gamma}}}

\def\Cset{\mathds{C}}
\def\Rset{\mathds{R}}

\def\vecalf{\boldsymbol{\alpha}}

\def\vecalfp{\boldsymbol{\alpha}^\prime}
\def\vecalfk{\boldsymbol{\alpha}_k}

\def\vecalfkp{\boldsymbol{\alpha}^\prime_k}
\def\vecD{{\bf D}}
\def\vecC{{\bf C}}

\begin{document}



\title{A Fluctuation-Dissipation Structure of Quantum Dynamical Semigroups Reveals a Unique Internal Hamiltonian}

\author{Fabricio  Toscano}   
\email{toscano@if.ufrj.br, \ corresponding author}
\affiliation{Instituto  de F\'isica,  Universidade  Federal do  Rio de  Janeiro,
  21941-972, Rio de Janeiro, Brazil}

\author{Sergey Sergeev}
\email{sergeevse1@im.ufrj.br}
\affiliation{Instituto de Matemática, Universidade  Federal do  Rio de  Janeiro,
  21941-972, Rio de Janeiro, Brazil}

\date{\today}

\begin{abstract}
We refine a fluctuation-dissipation framework for quantum dynamical semigroups to resolve a long-standing ambiguity in Markovian master equations. For finite-dimensional systems, we prove that the underlying diffusion-dissipation structure—rooted in a classical Markov process analogy—is invariant under Lindblad generator symmetries. This invariance uniquely identifies the internal Hamiltonian. Our framework provides a universal classification of quantum dynamical semigroups where the coherent and incoherent parts of the dynamics are uniquely determined thus enabling an unambiguous determination of a system's inherent energy structure.
\end{abstract}

\maketitle

{\it Introduction.} There is no physical system that can be considered completely isolated from its environment. The interaction between a system and its surroundings is the main mechanism that degrades quantum properties, which are essential for the development of quantum technologies.
Understanding these processes in order to mitigate their effects is currently the main objective of the theory of open quantum systems~\cite{Harrington2022}.
\par
When memory effects can be disregarded, the standard framework to describe the evolution is provided by quantum dynamical semigroups (QDS), where the system's density operator $\hrho_t$ obeys a Markovian master equation,
\begin{equation}
\label{eq:Markmastereq}
\frac{d\hrho_t}{dt}=\calL(\hrho_t),
\end{equation}
governed by a time-independent generator $\calL$~\cite{Alicki2007,Attal2006,Breuer2002,Fagnola2003,Manzano2020}.
\par
QDSs are universally applied across quantum physics, including quantum optics~\cite{Carmichael1998,Gardiner2004}, quantum information~\cite{Harrington2022,Rivas2012}, quantum communication~\cite{Heinosaari2010,Toscano2021}, quantum Brownian motion~\cite{Demoen1979,Heinosaari2010,Tarasov2008}, and quantum thermodynamics~\cite{Alicki2007,Carlen2017,Toscano2021,gemmer2009,Binder2019}, among many other applications.
\par
The general form of the infinitesimal generator is
\begin{equation}
\label{eq:decompUNU}
\calL = \calL_{\mathrm{U}} + \calL_{\mathrm{NU}},
\end{equation}
consisting of a unitary part $\calL_{\mathrm{U}}$, associated with a Hamiltonian $\hH$, and a non-unitary part $\calL_{\mathrm{NU}}$, governed by a set of Lindblad operators $\{\hL_k\}$.
This rigorous formulation was established first for finite-dimensional systems by Gorini, Kossakowski, and Sudarshan~\cite{Gorini1976}, and subsequently extended independently by Lindblad~\cite{Lindblad1976} to any $W^*$-algebra of operators.
\par
The universal applicability of the Lindblad master equation relies on the fact that it can be derived from the total system's unitary evolution under certain approximations. These approximations are: the weak-coupling (Born approximation), the memoryless reservoir (Markov approximation), and the separation of timescales (secular or rotating-wave approximation)~\cite{Davies1976,Breuer2002,Schlosshauer2007}. Furthermore, phenomenological approaches that preserve the Lindblad form are also common~\cite{Alicki2007}.
\par
A fundamental limitation of the Lindblad framework is the non-unique decomposition in Eq.~\eqref{eq:decompUNU}, which originates from a symmetry that transforms $\calL_{\mathrm{U}}$ and $\calL_{\mathrm{NU}}$ while leaving the total generator $\calL$ invariant~\cite{Lindblad1976,Breuer2002}. 
This ambiguity in identifying the internal Hamiltonian obstructs a unique definition of internal energy
representing  the primary obstacle for developing a consistent quantum thermodynamics from master equations~\cite{Levy2014,Kosloff2013,DeChiara2018,Binder2019,Landi2020}. Consequently, previous resolutions have relied on external principles, such as minimal dissipation~\cite{Colla2022,Hayden2022}.
 \par
 The decomposition~\eqref{eq:decompUNU} of the generator $\calL$ is defined modulo a symmetry group action, which induces equivalence classes 
of decompositions. 
The equivalence relation reflects the fact that all decompositions within a given class yield identical dynamical behavior.
We find that each element of the quotient space~\footnote{Remember that a quotient space is the space where each element is one class of equivalence.}, characterized by a given $\calL$, is also characterized by a unique decomposition
$\calL= \calL^\prime_{\mathrm{U}}+ \calL^\prime_{\mathrm{NU}}$. Thus, 
$\calL^\prime_{\mathrm{U}}$ and $\calL^\prime_{\mathrm{NU}}$ are invariant under the symmetry group action that conects the elements of 
each equivalence class.
\par
The unique decomposition that characterizes each element of the quotient space is obtained through 
a refinement of the fluctuation-dissipation framework for finite-dimensional QDSs introduced in~\cite{Toscano2021}. This framework cleanly separates the evolution into individually symmetry group invariant noise (fluctuation) and dissipation, associated with the superoperators $\calL_{\mathrm{fluc}}$ and $\calL_{\mathrm{diss}}$, respectively. The invariant infinitesimal generator for non-coherent processes is thus $\calL^\prime_{\mathrm{NU}}=\calL_{\mathrm{fluc}}+\calL_{\mathrm{diss}}$. Consequently, the invariant superoperator $\calL^\prime_{\mathrm{U}}$, for which $\calL= \calL^\prime_{\mathrm{U}}+ \calL^\prime_{\mathrm{NU}}$, governs the coherent evolution and is characterized by a unique Hamiltonian $\hHp$, challenging the common belief.
\par
The fluctuation-dissipation framework for finite-dimensional QDSs provides an alternative to the Lindblad formalism, based on a direct analogy with classical Markov processes described by linear Fokker-Planck equations. The invariance of $\calL_{\mathrm{fluc}}$ reflects the shared structure of quantum Langevin forces (noise) common to an infinite number of microscopic processes described by Lindblad operators. The same holds for $\calL_{\mathrm{diss}}$ in relation to dissipative effects. These invariant structures are encapsulated in two matrices: the diffusion matrix $\mD$ and the dissipation matrix $\mC$, which characterize $\calL_{\mathrm{fluc}}$ and $\calL_{\mathrm{diss}}$, respectively and thus the elements of the quotient space of decomposition classes in QDSs.
\par
Rooted in an analogy with classical Markov processes, an invariant diffusion-dissipation structure, and a unique internal energy, this new insight into QDS dynamics finds broad application. It advances quantum thermodynamics within the master equation formalism by resolving the principal obstacle to its development, \ie reavealing the unique internal energy structure, and by leveraging insights from the classical analogy.
Furthermore, it enables a systematic classification of stationary states in high-dimensional systems~\cite{Frigerio1977,Baumgartner2012,Albert2014} from the structure of the dissipation and diffusion, with direct applications in many-body systems~\cite{Olmos2012,Manzano2016,Mortimer2025,Lau2023,Mi2024,GuoHartChen2025,SouzaIemini2024}.
The framework also provides a powerful tool for analyzing dissipation engineering~\cite{Harrington2022}, such as in Zeno effect scenarios where identifying a unique internal Hamiltonian is crucial~\cite{ZhangFan2015}. Centered on the invariant structure of diffusion and dissipation, this new perspective offers a fundamental approach to environmental effects in quantum technologies. Also, it serves as an ideal framework for analyzing the distinct roles of diffusion and dissipation to elucidate the quantum-classical transition~\cite{Toscano2005,Toscano2009,Ranard2025}.
\par
{\it The result.} 
\par
In the Schr\"odinger picture a QDS is a  completely positive (CP)
and trace-preserving (TP) one-parameter semigroup $\{\Lamb_t=e^{\calL t}|t\ge 0\}$, that maps density operators of a quantum systems into itself, \ie $\hrho_t=\Lamb_t(\hrho_0)$.
The so called Liouville superoperator $\calL$ in Eq.~\eqref{eq:decompUNU}, according to Lindblad's theorem~\cite{Lindblad1976},  is given by
\bse
\label{eq:defsupL}
\begin{align}
\calL_{\mathrm{U}}(\cdot)&=-\frac{\imath }{\hbar}[\hH,\cdot],
\label{eq:defsupLUad}\\
\calL_{\mathrm{NU}}(\cdot)&=\frac{1}{2\hbar}\sum_{k=1}^K\left(2\hL_k\cdot\hL^\dagger_k-\{\hL^\dagger_k\hL_k,\cdot\}\right).
\label{eq:defsupLNUb}
\end{align}
\ese
\par
 The decomposition~\eqref{eq:decompUNU} of the Liouville superoperator in unitary and non-unitary contributions is not unique ~\cite{Breuer2002}. 
Indeed, $\calL $ is invariant under  the unitary transformation of the Lindblad operators:
	\begin{equation}
		\label{eq:unitarytransformation}
		\hL_k \rightarrow \hL^\prime_k = \sum^K_{j=1} \mW_{k,j} \hL_j,
	\end{equation}
where $\mW$ is an arbitrary unitary matrix, and also  invariant   under  the inhomogeneous
transformations for any $\beta_k \in \Cset$ and $b\in \Rset$:
\begin{subequations}
	\label{eq:Invtransf}
	\begin{align}
		\hL_k &\rightarrow \hL_k + \beta_k \hid\;\;\;,\;\;\;
		\hH \rightarrow \hH +  \hH^{\prime\prime},
		\label{eq:Invtransfa}\\
		\hH^{\prime\prime} &=  \frac{1}{2\imath }\sum_{k=1}^K   (  \beta_k^*   \hL_k  -   \beta_k  \hL_k^\dag)+b \hid.
		\label{eq:Invtransfb}
	\end{align} 
\end{subequations}
The transformations~\eqref{eq:unitarytransformation} and~\eqref{eq:Invtransf} verify a group property, thus they constitute the symmetry group of $\calL$ in Eq.~\eqref{eq:decompUNU}~\cite{Breuer2002}.
This symmetry group induces an equivalence relation on the sets of operators: two sets $\{\hH_1, \{\hL_k\}_1\}$ and $\{\hH_2, \{\hL_k\}_2\}$  are equivalent if one can be obtained from the other by the group transformations~\eqref{eq:unitarytransformation} and~\eqref{eq:Invtransf}.
\par
An equivalent characterization of $\calL_{\mathrm{NU}}$ in Eqs.~\eqref{eq:defsupLNUb}, was given in~\cite{Toscano2021} in terms of real superoperators $\calL_j$, $j=1,2,3$,
so 
\begin{equation}
\calL_{\mathrm{NU}}(\cdot)=\calL_1(\cdot) +\calL_2(\cdot) +\calL_3(\cdot) 
\label{eq:L1L2L3}
\end{equation}
with 
\begin{subequations}
\label{eq:L1L2L3sepa}
\begin{align}
\calL_1(\cdot)   &=  -\frac{1}{2\hbar} \sum_{k=1}^K  ( [\hA_k  , [\hA_k  , \cdot]]  +
    [\hB_k , [\hB_k ,\cdot]]),
\label{eq:L1} \\
\label{eq:L2}
\calL_2(\cdot) & = \frac{1}{2\hbar}  \sum_{k=1}^K \frac12 \{ [\hL_k ,\hL_k^\dag]
, \cdot \}, \text{and} \\
\label{eq:L3}
\calL_3(\cdot)  &  =  \frac{1}{2\hbar}  \sum_{k=1}^K  (\hL_k  \cdot  \hL^\dag_k  -
\hL_k^\dag \cdot \hL_k),
\end{align}
\end{subequations}
where the Hermitian operators 
$\hA_k= \frac{1}{2} (\hL_k +  \hL_k^\dag)$ and 
$\hB_k=\frac1{2  \imath}(\hL_k -\hL_k^\dag)$,
are the cartesian decomposition of the Lindblad operators, \ie
$\hL_k=\hA_k+\imath \hB_k$. 
The superoperators $\calL_1$ and $\calL_2$ are selfadjoint~\footnote{Where the adjoint superoperator 
is defined with respect to the Hilbert-Schmidt  scalar product \ie $\langle\calLb(\hA),\hrho\rangle_{\rm HS}=\langle\hA,\calL(\hrho)\rangle_{\rm HS}$ with $\langle\hB,\hA\rangle_{\rm HS}=\Tr(\hB^\dagger\hA)$.}
and $\calL_3$  is antisymmetric, \ie
$\bar{\calL}_1 = \calL_1$, $\bar{\calL}_2 = \calL_2$ and
$\bar{\calL}_3 =- \calL_3$.
Also, $\calL_1$, $\calL_2$ and $\calL_3$ remain invariant under the transformation in~\eqref{eq:unitarytransformation} and 
transform as $\calL_1 \rightarrow  \calL_1$, $\calL_2 \rightarrow  \calL_2$ and 
\beq
\calL_3 \rightarrow    \calL_3   -    \frac{1}{\imath\hbar}[\hH^{\prime\prime},\cdot] ,
\eeq 
under the transformations~\eqref{eq:Invtransf}.
\par 
This framework naturally identifies the quantum Langevin forces $\sqrt{\hbar}\hA_k$ and $\sqrt{\hbar}\hB_k$, which describe the noise corresponding to the fluctuation part of the dynamics, $\calL_1$, and separates the dissipation part, which is essentially given by $\calL_2$ and $\calL_3$. 
This separation is corroborated by the role these superoperators play in the rate of change of the von Neumann entropy of the system~\cite{Toscano2021}, in complete analogy with the behavior of the rate of change of the Shannon entropy for the probability density distribution in classical Markovian processes described by linear Fokker-Planck equations.
\par 
From now on we consider the fluctuation-dissipation framework of QDSs in finite dimensional systems. 
It can be described using the orthogonal basis $\{\hT_\nu\}$ in the 
associated Hilbert space $\calHb={\cal B}({\cal H})$, with $\dim(\calH)=N$, of all bounded operators acting on $\calH$.
In the following greek letters subindexes run from $0,\ldots,N^2-1$ and italic subindices run from $1,\ldots,N^2-1$.
Operator $\hT_0=\sqrt{\frac{2}{N}}\hid$, where $\hid$ is the identity operator acting on $\calH$, so $\Tr(\hid)=N$,
and $\{\hT_i\}$ are  traceless Hermitian operators being a basis of the $\mathfrak{su}(N)$ Lie algebra such
\beq
\label{eq:TiTj}
\hT_i\hT_j=\frac{2}{N}\delta_{ij}\hid_\calH+\sum_l(\imath f_{ijl}+d_{ijl})\hT_l,
\eeq 
where $f$ is the totally antisymmetric tensor of structure constants and $d$ the totally symmetric one
\cite{Macfarlane1968}. The elements of $\{\hT_\nu\}$ satisfy the orthogonal condition
$\expval{\hT_\nu,\hT_\mu}=\Tr(\hT_\nu^\dagger\hT_\mu)=2\delta_{\nu\mu}$. The factor $2$ in the right hand side is a normalization that guarantee that for $N=2$ the operators $\{\hT_i\}_{i=1}^3$ are the set of Pauli operators and for $N=3$, $\{\hT_i\}_{i=1}^8$ is the set of Gell-Mann operators.
\par
We can write the Lindblad operators in the basis $\{\hT_\nu\}$:
\beq
\label{eq:Lk_expan}
\hL_k=\sum_\nu \alk_{\nu}\hT_\nu,
\eeq
with $\alk_\nu\in \Cset$. We define the positive semidefinite Hermitian matrix $\mGamt=\sum_{k=1}^K\vecalfk
\vecalfk^\dagger=\mA\mA^\dagger.$
Here $\mA=(\vecalf_1,\ldots,\vecalf_{K})$ is a $N^2\times K$ complex matrix
with the column vectors $\vecalfk=(\alk_0,\ldots,\alk_{N^2-1})^\tp$. 
We can write 
\beq
\label{eq:matr_Gamma}
\mGamt=\frac{\mDt}{\hbar}+\imath\mCt,
\eeq
where we define the real symmetric matrix $\mDt=\hbar \Re(\mGamt)$, and the real antisymmetric matrix $\mCt=\Im(\mGamt)$. It is important to highlight the following structure of these matrices:
\beq
\mDt= \begin{pmatrix} 
      \mDt_{00} & \vecD^\tp  \\
      \vecD & \mD 
   \end{pmatrix}\;\;,\;\;
\mCt= \begin{pmatrix} 
      0 & \vecC^\tp\\
    - \vecC& \mC
   \end{pmatrix}.
\label{eq:partsofGamt} 
\eeq
Here, we define the real vectors, $\vecD=(\mDt_{01},\ldots,\mDt_{0\,N^2-1})^\tp$ and $\vecC=(\mCt_{01},\ldots,\mCt_{0\,N^2-1})^\tp$, and  
the real $N^2-1\times N^2-1$ block matrices, $\mD$ and $\mC$, that are symmetric and antisymmetric respectively.
\par
We substitute~\eqref{eq:Lk_expan} into~\eqref{eq:L1L2L3sepa} and use~\eqref{eq:TiTj} with linearity of the  commutators and anti-commutators. Finally we use~\eqref{eq:matr_Gamma} and~\eqref{eq:partsofGamt}, which leads to (see the Appendix for details)
\bse
\label{eq:newL1L2L3decomp}
\begin{align}
\calL_1(\cdot)&=
-\frac{1}{2\hbar^2}\sum_i\sum_j
\mD_{ij}\;[\hT_i,[\hT_j,\cdot]],
\label{eq:L1dimfinite}\\
\calL_2(\cdot)&=\frac{1}{\hbar}\sum_i\sum_j\imath\mC_{ij}\frac{1}{2}\{\hT_i\hT_j,\cdot\}
\label{eq:L2dimfinitea}\\
&=\frac{1}{2\hbar}\sum_l\tr(\imath \mC \,\mf^l)\,\{\hT_l,\cdot\},
\label{eq:L2dimfinite}\\
\calL_3^\prime(\cdot)&=
\frac{1}{\hbar}
\sum_i\sum_j
\imath\,\mC_{ij}\hT_i\cdot\hT_j,
\label{eq:L3pdimfinite}
\end{align}
\ese
where $\calL_3(\cdot)=-\frac{\imath}{\hbar}\left[\hHb,\cdot\right]+\calL^\prime_3(\cdot)$,
with
\begin{align}
\hHb&=\sqrt{\frac{2}{N}}\sum_i(\vecC)_{i}\hT_i\nonumber\\
&=-\frac{1}{2\imath\,N }\sum_{k=1}^K   \left( \Tr(\hL_k)^*  \hL_k  -    \Tr(\hL_k)   \hL_k^\dag\right)
\label{eq:defHb}
\end{align}
with $(\vecC)_i$ the vector's coordinates and we used that $\Tr(\hL_k)=\alpha_0^{(k)}\sqrt{\frac{2}{N}}N$. The $N^2-1\times N^2-1$ Hermitian matrices $\{\mf^l\}$, whose 
matrix elements are determined by the structure constants, \ie 
$\mf^l_{i j}=-\imath f_{lij}$, are the generators of the $\mathfrak{su}(N)$ algebra in the adjoint representation~\cite{Haber2021}.
\par
Now,
we can present the fluctuation-dissipation structure of the Liouville superoperator $\calL$ of a finite-dimensional QDS as
\beq
\label{eq:mainres1}
\calL=\calL^\prime_{\mathrm{U}}+\calL^\prime_{\mathrm{NU}}.
\eeq
Here, we define
\bse
\begin{align}
\calL^\prime_{\mathrm{U}}(\cdot)&=-\frac{\imath}{\hbar}\left[\hHp,\cdot\right],
\label{eq:defcalLUp}\\
\hHp&=\hH+\hHb
\label{eq:defHint}
\end{align}
\ese
with $\hH$ the original Hamiltonian in~\eqref{eq:defsupLUad} and $\hHb$ defined in~\eqref{eq:defHb}.
Also, 
\bse
\label{eq:defLNUprime}
\begin{align}
\calL^\prime_{\mathrm{NU}}&=\calL_{\mathrm{fluc}}+\calL_{\mathrm{diss}},
\label{eq:defLNUprimea}\\
\calL_{\mathrm{fluc}}&=\calL_1\;\text{and}\; \calL_{\mathrm{diss}}=\calL_2+\calL_3^\prime.
\label{eq:defLNUprimeb}
\end{align}
\ese 
Interesting enough, according to~\eqref{eq:newL1L2L3decomp}, the evolution of a finite-dimensional QDS, does not depend on $\mDt_{00}$ and the vector $\vecD$, that are block sectors of the matrix $\mDt$ in~\eqref{eq:partsofGamt}.
\par
\par
A key point is that the fluctuation dynamics $\calL_{\mathrm{fluc}}$ is governed solely by the diffusion  block  matrix $\mD$, whereas the dissipation $\calL_{\mathrm{diss}}$ is determined by the dissipation  block matrix $\mC$.
Because $\hbar\mGamt \ge 0$ and $\mDt_{00}=\sum_{k=1}^K|\alpha_0^{(k)}|^2\ge 0$, we have $\mD+\imath \hbar \mC \ge 0$. 
So, $\mD$ can not be the null matrix unless $\mC$ is also the null matrix. 
Thus, while QDSs without dissipation, \ie $\mC=0$, is possible, there are no QDSs without diffussion, \ie with $\mD=0$.
If  $N^2-1$ is even and  $\det(\mC)\neq 0$, then both $\mC$ and $\mD$ are invertible
and $\mD>0$.  If  $N^2-1$ is odd, we always have $\det(\mC)= 0$, so generically $\mD\ge 0$ in this case. 
\par
{\sl Theorem  (Invariance of the Fluctuation-Dissipation Structure). Under the group of symmetry transformations of QDSs given by Eqs.~\eqref{eq:unitarytransformation} and~\eqref{eq:Invtransf}, the superoperators defined in Eqs.~\eqref{eq:newL1L2L3decomp} are individually invariant. Therefore, $\calL_{\mathrm{fluc}}$ and $\calL_{\mathrm{diss}}$ in~\eqref{eq:defLNUprimeb} are invariant, and consequently so is $\calL^\prime_{\mathrm{NU}}$ 
in~\eqref{eq:defLNUprimea}}.\\
While the invariance of $\calLp_{\mathrm{U}}$ follows from the invariance of $\calL$ and the decomposition in~\eqref{eq:mainres1}, we also provide an independent demonstration of this fact to corroborate the consistency of our approach.\\
 {\sl Proof:} First note that under the transformation in~\eqref{eq:unitarytransformation} the matrix $\mGamt$ is invariant.
 Indeed,  we have the new $\mGamt^\prime=\mAp(\mAp)^\dagger$, where  $\mAp=(\vecalfp_1,\ldots,\vecalfp_{K})=\mathbb{A}\mW^\tp$,
 using the vectors $\vecalfkp$ associated with $\hLp_k$ in the basis $\{\hT_\nu\}$, and $\mW$ is the unitary matrix in~\eqref{eq:unitarytransformation}.
So, we have $\mGamt^\prime=\mathbb{A}'\mathbb{A}^{\prime\dagger}=\mathbb{A}\mW^\tp(\mW^\tp)^\dagger\mathbb{A}^\dagger=\mGamt$
and consequently the diffusion and dissipation matrices, $\mD$ and $\mC$ respectively, are invariant under the transformation in~\eqref{eq:unitarytransformation}.  
\par
Now, applying the  transformation in~\eqref{eq:Invtransfa}, the new Lindblad operators are 
$ \hL_k + \beta_k \hid=\left( \sqrt{\frac{N}{2}}\beta_k+\alk_{0}\right)\hT_0+\sum_i \alk_{i}\hT_i$.
Therefore, the only blocks sectors of $\mGamt$, in~\eqref{eq:matr_Gamma} and~\eqref{eq:partsofGamt}, that transform are:
$\frac{\mDt_{00}}{\hbar}\rightarrow
\frac{\mDt_{00}}{\hbar}+\sum_{k=1}^K \frac{N}{2}|\beta_k|^2+2\sum_{k=1}^K \sqrt{\frac{N}{2}}\Re(\beta_k\alkb_0)$,
$\vecD\rightarrow  \vecD+\vecD^\prime$, with
$\frac{(\vecD^\prime)_{0i}}{\hbar}=\sum_{k=1}^K \sqrt{\frac{N}{2}}\Re\left(\beta_k\alkb_i\right)$ 
and
\bse
\label{eq:Gamttransformed}
\begin{align}
\vecC&\rightarrow  \vecC+\vecC^\prime,\\
(\vecC^\prime)_i&=\sum_{k=1}^K \sqrt{\frac{N}{2}}\Im\left(\beta_k\alkb_i\right).
\end{align}
\ese
Therefore, the block sectors $\mD$ and $\mC$ in~\eqref{eq:partsofGamt} remain invariants.
\par
Thus, we have shown that the diffusion and dissipation matrices, $\mD$ and $\mC$ respectively, are invariant under the transformations in~\eqref{eq:unitarytransformation} and~\eqref{eq:Invtransfa}, and because 
the superoperators $\calL_1$, $\calL_2$ and $\calL_3^\prime$ in Eqs.\eqref{eq:newL1L2L3decomp}, only depend on these
matrices, they are independently invariant under these transformations.
\par
Let's now corroborate that $\calL^\prime_{\mathrm{U}}$ in~\eqref{eq:mainres1} is invariant under the transformation in~\eqref{eq:Invtransfa}. For this it is sufficient to prove that $\hHp$ in~\eqref{eq:defHint} is invariant.
This is true because under this transformation, we have $\Tr(\hL_k+\beta_k\hid)=\Tr(\hL_k)+\beta_k \,N$, so $\hHb\rightarrow \hHb- \hH^{\prime\prime}$ while $\hH\rightarrow \hH + \hH^{\prime\prime}$ with $ \hH^{\prime\prime}$ given in Eq.~\eqref{eq:Invtransfb}, so the transformed Hamiltonian $\hHp=\hH + \hH^{\prime\prime}+\hHb - \hH^{\prime\prime}$ remains the same.
QED. 
\par
Therefore, if a system's dynamics description—from heuristic or first-principles evaluation—uses Lindblad operators with non-zero trace, the unique Hamiltonian $\hHp$ in~\eqref{eq:defHint} may contain a non-zero $\hHb$. The theorem asserts, however, that $\calL^\prime_{\mathrm{U}}$ and $\calL^\prime_{\mathrm{NU}}$ remain invariant for any other set of Lindblad operators (with any trace) satisfying Eqs.~\eqref{eq:unitarytransformation} or~\eqref{eq:Invtransfa} that describe the same system's dynamics.
\par
The change to a new operator basis in~\eqref{eq:Lk_expan}, where both bases satisfy Eq.~\eqref{eq:TiTj}, is performed by an orthogonal matrix $\mO$ via $\hTp_i=\sum_j (\mO^\tp)_{ij} \hT_j$. This transforms the matrices as $\mDp=\mO^\tp \mD \mO$ and $\mCp= \mO^\tp \mC \mO$. Substituting into Eqs.~\eqref{eq:L1dimfinite},~\eqref{eq:L2dimfinitea} and~\eqref{eq:L3pdimfinite} and using the orthogonality of $\mO$, we confirm the invariance of $\calL_1$, $\calL_2$, and $\calLp_3$. Thus, $\calL_{\mathrm{fluc}}$ and $\calL_{\mathrm{diss}}$ in~\eqref{eq:defLNUprime} are also invariant under the change of operator basis.
\par
{\sl Corollary. If a QDS has a $\hH$ from~\eqref{eq:defsupLUad} and a traceless set of Lindblad operators $\{\hL_k\}$ from~\eqref{eq:defsupLNUb}, then $\calL$ is already in the invariance form of~\eqref{eq:mainres1} with $\hHp = \hH$, $\calL_{\mathrm{U}}=\calLp_{\mathrm{U}}$ and $\calL_{\mathrm{NU}}=\calLp_{\mathrm{NU}}$.}
\\
 {\sl Proof:} Note that $\calL_1$ and $\calL_2$ are invariant, and from~\eqref{eq:defHb} we have $\hHb=0$, so $\calL_3=\calLp_3$ is also invariant. QED. 
\par
Note that corollary does not claim the existence of a unique pair $\calL_{\mathrm{U}}$ and $\calL_{\mathrm{NU}}$—where the former is generated by a Hamiltonian with ${\rm Tr}(\hH)=0$ and the latter by a set of traceless operators $\{\hL_k\}$—such that $\calL=\calL_{\mathrm{U}}+\calL_{\mathrm{NU}}$, as demonstrated in~\cite{Gorini1976}. 
As shown in~\cite{Hayden2022, vomEnde2024}, superoperators associated with traceless Lindblad operators realize the unique orthogonal decomposition of the vector space of Liouville superoperators under certain inner products. This approach represents the entire vector space as an orthogonal direct sum of two subspaces, corresponding respectively to the Hamiltonian and non-unitary parts of the Liouville superoperator. Consequently, within any equivalence class of decompositions of the form in Eq.~\eqref{eq:decompUNU}, the decomposition associated with traceless operators minimizes the norm $\|\calL_{\mathrm{NU}}\|$ with respect to the defined inner products. This provides the justification for the so-called principle of minimum dissipation~\cite{Hayden2022, Colla2022}, which selects the Hamiltonian associated with traceless Lindblad operators as the internal energy of the system.
\par
Our approach removes this arbitrariness: the Corollary shows that all possible decompositions $\{\hat{H}, \{\hat{L}_k\}\}$ that generate the same physical evolution collapse onto a single, unique fluctuation-dissipation structure, precisely the one given by these traceless Lindblad operators and their corresponding unique internal energy.
\par
For a finite-dimensional system with a general system-environment interaction $\hH_I=\sum_{i} \hT_i\otimes \hB_i$, the Born-Markov and secular approximations yield a Liouvillian $\calL$ described  by traceless Lindblad operators $\hL_k=\sum_{i} c_{ki}(\omega) \hT_i(\omega)$ and a Hamiltonian $\hH_S+\hH_{LS}$, with the Lamb shift $\hH_{LS}=\sum_{\omega,i, j} S_{ij}(\omega) \hT_i^\dagger(\omega) \hT_{j}(\omega)$ commuting with $\hH_S$. These are constructed from the system's energy eigenoperators $\hT_{i}(\omega) = \sum_{\epsilon_a - \epsilon_b = \omega} \dyad{\epsilon_a} \hT_{i} \dyad{\epsilon_b}$, where $\hH_S\ket{\epsilon_a}=\epsilon_a\ket{\epsilon_a}$. Thus, applying the corollary the unique Hamiltonian 
is $\hHp=\hH_S+\hH_{LS}$.
\par 
{\it Final remarks.}
A primary objective of open quantum systems theory is to define an internal energy for arbitrary interacting systems. For quantum dynamical semigroups (QDSs), this definition has long been regarded as intrinsically ambiguous, with resolutions typically relying on external principles such as minimal dissipation~\cite{Colla2022, Hayden2022}. In this work, we resolve this ambiguity by employing a fluctuation-dissipation framework that uniquely identifies the Hamiltonian governing the unitary evolution, thereby enabling a physically consistent definition of internal energy.
\par
We introduce a framework for finite-dimensional QDSs that characterizes the non-unitary part $\calLp_{\mathrm{NU}}$ in~\eqref{eq:defLNUprimea} via two real matrices: a symmetric diffusion matrix $\mD$, which embodies quantum Langevin forces, and an antisymmetric dissipation matrix $\mC$, which encodes dissipative effects. This formulation is expressed directly in terms of these physical matrices rather than Lindblad operators. We prove that $\mD$ and $\mC$ are invariant under the full symmetry group of the generator $\calL$.
\par
Consequently, all equivalent decompositions of the Liouville superoperator collapse into the form given in Eq.~\eqref{eq:mainres1}, which depends only on these invariant matrices. This allows the well-known detailed balance condition~\cite{SpohnLebowitz1978, Alicki1976}—which ensures thermal equilibrium—to be formulated in an invariant and consistent manner entirely in terms of $\mD$ and $\mC$. Hence, the thermal equilibrium state is the Gibbs state $\hat{\rho}_{\mathrm{G}} = e^{-\beta \hat H'}/\mathcal{Z}$, with the unique Hamiltonian $\hat H'$ given in Eq.~\eqref{eq:defcalLUp}~\footnote{Details will be published elsewhere.}.
\par
Many distinct physical processes can yield identical fluctuation noise ($\mD$) and dissipation ($\mC$), thereby producing the same dynamics. For example, a qubit with $\mD=\frac{\gamma}{2}\operatorname{diag}(1,1,0)$ and $\mC=0$ can represent either an amplitude damping channel with an infinite-temperature bath (described by Lindblad operators $\hL_1 = \sqrt{\Gamma n_{\text{th}}}\,\hsig_+$ and $\hL_2 = \sqrt{\Gamma(n_{\text{th}}+1)}\,\hsig_-$, where $n_{\text{th}}\Gamma \rightarrow \gamma$, $\hsig_{\pm}=\frac{1}{2}(\hsig_1\pm i \hsig_2)$, and $\hsig_i$ are the Pauli operators) or a depolarizing channel (described by $\hLp_i=\sqrt{\frac{\gamma}{2}}\,\hsig_i$ for $i=1,2$). The sets $\{\hL_i\}$ and $\{\hLp_i\}$ are connected by the symmetry transformation in~\eqref{eq:unitarytransformation}, which leaves $\mD$ invariant, as well as by those in Eq.~\eqref{eq:Invtransf}.
\par
From Eq.~\eqref{eq:L2}, all quantum Markov semigroups (QMSs)—i.e., unital QDSs in the Schrödinger picture—must satisfy $\calL_2\equiv 0$ [48]. Thus, Eq.~\eqref{eq:L2dimfinite} reveals two distinct types of QMSs: {\it i)} those with $\mC\neq 0$ and $\tr(\imath \mC ,\mf^l)=0$, and {\it ii)} those with $\mC=0$. For qubit systems, only case {\it ii)} is possible, so the invariant non-unitary contribution to the generator reduces to $\calLp_{\mathrm{NU}}=\calL_{\mathrm{fluc}}$. This classification becomes considerably more intricate when expressed in terms of Lindblad operators.

Finally, since the infinitesimal generator of any time-local master equation admits a Lindblad form with an associated symmetry group~\cite{Hall2014} (see also~\cite{Hayden2022,Colla2022}), our results can be extended to both Markovian and non-Markovian dynamics [48].
It is worth noting that in finite dimensions, the exact dynamics of the reduced state of two arbitrary interacting systems is described by a time-local master equation except on a set of isolated points in time~\cite{Hayden2022}. Therefore, we believe, the fluctuation-dissipation approach will constitute a significant contribution to the foundations of quantum thermodynamics for arbitrary system-environment interactions.

\acknowledgments{S.S. thanks FAPERJ project APQ1 E-26/210.614/2024.}


\bibliographystyle{apsrev4-2}

\begin{thebibliography}{49}%
\makeatletter
\providecommand \@ifxundefined [1]{%
 \@ifx{#1\undefined}
}%
\providecommand \@ifnum [1]{%
 \ifnum #1\expandafter \@firstoftwo
 \else \expandafter \@secondoftwo
 \fi
}%
\providecommand \@ifx [1]{%
 \ifx #1\expandafter \@firstoftwo
 \else \expandafter \@secondoftwo
 \fi
}%
\providecommand \natexlab [1]{#1}%
\providecommand \enquote  [1]{``#1''}%
\providecommand \bibnamefont  [1]{#1}%
\providecommand \bibfnamefont [1]{#1}%
\providecommand \citenamefont [1]{#1}%
\providecommand \href@noop [0]{\@secondoftwo}%
\providecommand \href [0]{\begingroup \@sanitize@url \@href}%
\providecommand \@href[1]{\@@startlink{#1}\@@href}%
\providecommand \@@href[1]{\endgroup#1\@@endlink}%
\providecommand \@sanitize@url [0]{\catcode `\\12\catcode `\$12\catcode
  `\&12\catcode `\#12\catcode `\^12\catcode `\_12\catcode `\%12\relax}%
\providecommand \@@startlink[1]{}%
\providecommand \@@endlink[0]{}%
\providecommand \url  [0]{\begingroup\@sanitize@url \@url }%
\providecommand \@url [1]{\endgroup\@href {#1}{\urlprefix }}%
\providecommand \urlprefix  [0]{URL }%
\providecommand \Eprint [0]{\href }%
\providecommand \doibase [0]{https://doi.org/}%
\providecommand \selectlanguage [0]{\@gobble}%
\providecommand \bibinfo  [0]{\@secondoftwo}%
\providecommand \bibfield  [0]{\@secondoftwo}%
\providecommand \translation [1]{[#1]}%
\providecommand \BibitemOpen [0]{}%
\providecommand \bibitemStop [0]{}%
\providecommand \bibitemNoStop [0]{.\EOS\space}%
\providecommand \EOS [0]{\spacefactor3000\relax}%
\providecommand \BibitemShut  [1]{\csname bibitem#1\endcsname}%
\let\auto@bib@innerbib\@empty
\bibitem [{\citenamefont {Harrington}\  {et~al.}(2022)\citenamefont
  {Harrington}, \citenamefont {Mueller},\ and\ \citenamefont
  {Murch}}]{Harrington2022}%
  \BibitemOpen
  \bibfield  {author} {\bibinfo {author} {\bibfnamefont {P.~M.}\ \bibnamefont
  {Harrington}}, \bibinfo {author} {\bibfnamefont {E.~J.}\ \bibnamefont
  {Mueller}},\ and\ \bibinfo {author} {\bibfnamefont {K.~W.}\ \bibnamefont
  {Murch}},\ }\href {https://doi.org/10.1038/s42254-022-00494-8} {\bibfield
  {journal} {\bibinfo  {journal} {Nature Reviews Physics}\ }\textbf {\bibinfo
  {volume} {4}},\ \bibinfo {pages} {660–671} (\bibinfo {year}
  {2022})}\BibitemShut {NoStop}%
\bibitem [{\citenamefont {Alicki}\ and\ \citenamefont
  {Lendi}(2007)}]{Alicki2007}%
  \BibitemOpen
  \bibfield  {author} {\bibinfo {author} {\bibfnamefont {R.}~\bibnamefont
  {Alicki}}\ and\ \bibinfo {author} {\bibfnamefont {K.}~\bibnamefont {Lendi}},\
  }\href {https://books.google.com.br/books?id=1F_ZhntUqZsC} { {\bibinfo
  {title} {Quantum Dynamical Semigroups and Applications}}},\ Lecture Notes in
  Physics\ (\bibinfo  {publisher} {Springer Berlin Heidelberg},\ \bibinfo
  {year} {2007})\BibitemShut {NoStop}%
\bibitem [{\citenamefont {Stéphane~Attal}(2006)}]{Attal2006}%
  \BibitemOpen
  \bibfield  {author} {\bibinfo {author} {\bibfnamefont {C.-A.~P.}\
  \bibnamefont {Stéphane~Attal}, \bibfnamefont {Alain~Joye}},\ }\href
  {https://doi.org/10.1007/b128451} { {\bibinfo {title} {Open Quantum
  Systems II: The Markovian Approach}}},\ \bibinfo {series} {Lecture Notes in
  Mathematics}, Vol.\ \bibinfo {volume} {1881}\ (\bibinfo  {publisher}
  {Springer Berlin Heidelberg},\ \bibinfo {year} {2006})\BibitemShut {NoStop}%
\bibitem [{\citenamefont {Breuer}\ and\ \citenamefont
  {Petruccione}(2002)}]{Breuer2002}%
  \BibitemOpen
  \bibfield  {author} {\bibinfo {author} {\bibfnamefont {H.}~\bibnamefont
  {Breuer}}\ and\ \bibinfo {author} {\bibfnamefont {F.}~\bibnamefont
  {Petruccione}},\ }\href {https://books.google.com.br/books?id=0Yx5VzaMYm8C}
  { {\bibinfo {title} {The Theory of Open Quantum Systems}}}\ (\bibinfo
  {publisher} {Oxford University Pre},\ \bibinfo {year} {2002})\BibitemShut
  {NoStop}%
\bibitem [{\citenamefont {Fagnola}\ and\ \citenamefont
  {Rebolledo}(2003)}]{Fagnola2003}%
  \BibitemOpen
  \bibfield  {author} {\bibinfo {author} {\bibfnamefont {F.}~\bibnamefont
  {Fagnola}}\ and\ \bibinfo {author} {\bibfnamefont {R.}~\bibnamefont
  {Rebolledo}},\ }in\ \href {https://doi.org/10.1007/978-3-0348-8018-3_6}
  { {\bibinfo {booktitle} {Stochastic Analysis and Mathematical Physics
  II}}},\ \bibinfo {editor} {edited by\ \bibinfo {editor} {\bibfnamefont
  {R.}~\bibnamefont {Rebolledo}}}\ (\bibinfo  {publisher} {Birkh\"{a}user
  Basel},\ \bibinfo {year} {2003})\ pp.\ \bibinfo {pages} {77--128}\BibitemShut
  {NoStop}%
\bibitem [{\citenamefont {Manzano}(2020)}]{Manzano2020}%
  \BibitemOpen
  \bibfield  {author} {\bibinfo {author} {\bibfnamefont {D.}~\bibnamefont
  {Manzano}},\ }\href {https://doi.org/10.1063/1.5115323} {\bibfield  {journal}
  {\bibinfo  {journal} {AIP Advances}\ }\textbf {\bibinfo {volume} {10}},\
  \bibinfo {pages} {025106} (\bibinfo {year} {2020})}\BibitemShut {NoStop}%
\bibitem [{\citenamefont {Carmichael}(1998)}]{Carmichael1998}%
  \BibitemOpen
  \bibfield  {author} {\bibinfo {author} {\bibfnamefont {H.}~\bibnamefont
  {Carmichael}},\ }\href {https://books.google.com.br/books?id=ocgRgM-yJacC}
  { {\bibinfo {title} {Statistical Methods in Quantum Optics 1: Master
  Equations and Fokker-Planck Equations}}},\ Physics and astronomy online
  library\ (\bibinfo  {publisher} {Springer},\ \bibinfo {year}
  {1998})\BibitemShut {NoStop}%
\bibitem [{\citenamefont {Gardiner}(2004)}]{Gardiner2004}%
  \BibitemOpen
  \bibfield  {author} {\bibinfo {author} {\bibfnamefont {C.}~\bibnamefont
  {Gardiner}},\ }\href {https://books.google.com.br/books?id=wLm7QgAACAAJ}
  { {\bibinfo {title} {Handbook of Stochastic Methods for Physics,
  Chemistry, and the Natural Sciences}}},\ Springer complexity\ (\bibinfo
  {publisher} {Springer},\ \bibinfo {year} {2004})\BibitemShut {NoStop}%
\bibitem [{\citenamefont {Rivas}\ and\ \citenamefont
  {Huelga}(2011)}]{Rivas2012}%
  \BibitemOpen
  \bibfield  {author} {\bibinfo {author} {\bibfnamefont {{\'A}.}~\bibnamefont
  {Rivas}}\ and\ \bibinfo {author} {\bibfnamefont {S.}~\bibnamefont {Huelga}},\
  }\href {https://books.google.com.br/books?id=FGCuYsIZAA0C} { {\bibinfo
  {title} {Open Quantum Systems: An Introduction}}},\ SpringerBriefs in
  Physics\ (\bibinfo  {publisher} {Springer Berlin Heidelberg},\ \bibinfo
  {year} {2011})\BibitemShut {NoStop}%
\bibitem [{\citenamefont {Heinosaari}\  {et~al.}(2010)\citenamefont
  {Heinosaari}, \citenamefont {Holevo},\ and\ \citenamefont
  {Wolf}}]{Heinosaari2010}%
  \BibitemOpen
  \bibfield  {author} {\bibinfo {author} {\bibfnamefont {T.}~\bibnamefont
  {Heinosaari}}, \bibinfo {author} {\bibfnamefont {A.}~\bibnamefont {Holevo}},\
  and\ \bibinfo {author} {\bibfnamefont {M.}~\bibnamefont {Wolf}},\ }\href
  {https://doi.org/10.26421/QIC10.7-8-4} {\bibfield  {journal} {\bibinfo
  {journal} {Quantum Information and Computation}\ }\textbf {\bibinfo {volume}
  {10}},\ \bibinfo {pages} {619} (\bibinfo {year} {2010})}\BibitemShut
  {NoStop}%
\bibitem [{\citenamefont {Toscano}\  {et~al.}(2021)\citenamefont
  {Toscano}, \citenamefont {Bosyk}, \citenamefont {Zozor},\ and\ \citenamefont
  {Portesi}}]{Toscano2021}%
  \BibitemOpen
  \bibfield  {author} {\bibinfo {author} {\bibfnamefont {F.}~\bibnamefont
  {Toscano}}, \bibinfo {author} {\bibfnamefont {G.~M.}\ \bibnamefont {Bosyk}},
  \bibinfo {author} {\bibfnamefont {S.}~\bibnamefont {Zozor}},\ and\ \bibinfo
  {author} {\bibfnamefont {M.}~\bibnamefont {Portesi}},\ }\href
  {https://doi.org/10.1103/PhysRevA.104.062207} {\bibfield  {journal} {\bibinfo
   {journal} {Physical Review A}\ }\textbf {\bibinfo {volume} {104}},\ \bibinfo
  {pages} {062207} (\bibinfo {year} {2021})}\BibitemShut {NoStop}%
\bibitem [{\citenamefont {Demoen}\  {et~al.}(1979)\citenamefont {Demoen},
  \citenamefont {Vanheuverzwijn},\ and\ \citenamefont {Verbeure}}]{Demoen1979}%
  \BibitemOpen
  \bibfield  {author} {\bibinfo {author} {\bibfnamefont {B.}~\bibnamefont
  {Demoen}}, \bibinfo {author} {\bibfnamefont {P.}~\bibnamefont
  {Vanheuverzwijn}},\ and\ \bibinfo {author} {\bibfnamefont {A.}~\bibnamefont
  {Verbeure}},\ }\href {https://doi.org/10.1016/0034-4877(79)90049-1}
  {\bibfield  {journal} {\bibinfo  {journal} {Reports on Mathematical Physics}\
  }\textbf {\bibinfo {volume} {15}},\ \bibinfo {pages} {27–39} (\bibinfo
  {year} {1979})}\BibitemShut {NoStop}%
\bibitem [{\citenamefont {Tarasov}(2008)}]{Tarasov2008}%
  \BibitemOpen
  \bibfield  {author} {\bibinfo {author} {\bibfnamefont {V.}~\bibnamefont
  {Tarasov}},\ }\href {https://doi.org/true} { {\bibinfo {title} {Quantum
  mechanics of non-hamiltonian and diipative systems}}}\ (\bibinfo  {publisher}
  {Elsevier},\ \bibinfo {year} {2008})\BibitemShut {NoStop}%
\bibitem [{\citenamefont {Carlen}\ and\ \citenamefont
  {Maas}(2017)}]{Carlen2017}%
  \BibitemOpen
  \bibfield  {author} {\bibinfo {author} {\bibfnamefont {E.~A.}\ \bibnamefont
  {Carlen}}\ and\ \bibinfo {author} {\bibfnamefont {J.}~\bibnamefont {Maas}},\
  }\href {https://doi.org/https://doi.org/10.1016/j.jfa.2017.05.003} {\bibfield
   {journal} {\bibinfo  {journal} {Journal of Functional Analysis}\ }\textbf
  {\bibinfo {volume} {273}},\ \bibinfo {pages} {1810} (\bibinfo {year}
  {2017})}\BibitemShut {NoStop}%
\bibitem [{\citenamefont {Gemmer}\  {et~al.}(2009)\citenamefont {Gemmer},
  \citenamefont {Michel},\ and\ \citenamefont {Mahler}}]{gemmer2009}%
  \BibitemOpen
  \bibfield  {author} {\bibinfo {author} {\bibfnamefont {J.}~\bibnamefont
  {Gemmer}}, \bibinfo {author} {\bibfnamefont {M.}~\bibnamefont {Michel}},\
  and\ \bibinfo {author} {\bibfnamefont {G.}~\bibnamefont {Mahler}},\ }\href
  {https://doi.org/10.1007/978-3-540-70510-9} { {\bibinfo {title} {Quantum
  thermodynamics}}},\ Vol.\ \bibinfo {volume} {784}\ (\bibinfo  {publisher}
  {Springer},\ \bibinfo {year} {2009})\BibitemShut {NoStop}%
\bibitem [{\citenamefont {Binder}\  {et~al.}(2019)\citenamefont {Binder},
  \citenamefont {Correa}, \citenamefont {Gogolin}, \citenamefont {Anders},\
  and\ \citenamefont {Adeo}}]{Binder2019}%
  \BibitemOpen
  \bibinfo {editor} {\bibfnamefont {F.}~\bibnamefont {Binder}}, \bibinfo
  {editor} {\bibfnamefont {L.~A.}\ \bibnamefont {Correa}}, \bibinfo {editor}
  {\bibfnamefont {C.}~\bibnamefont {Gogolin}}, \bibinfo {editor} {\bibfnamefont
  {J.}~\bibnamefont {Anders}},\ and\ \bibinfo {editor} {\bibfnamefont
  {G.}~\bibnamefont {Adeo}},\ eds.,\ \href@noop {} { {\bibinfo {title}
  {Thermodynamics in the Quantum Regime: Fundamental Aspects and New
  Directions}}},\ \bibinfo {series} {Fundamental Theories of Physics}, Vol.\
  \bibinfo {volume} {195}\ (\bibinfo  {publisher} {Springer International
  Publishing},\ \bibinfo {year} {2019})\BibitemShut {NoStop}%
\bibitem [{\citenamefont {Gorini}\  {et~al.}(1976)\citenamefont {Gorini},
  \citenamefont {Koakowski},\ and\ \citenamefont {Sudarshan}}]{Gorini1976}%
  \BibitemOpen
  \bibfield  {author} {\bibinfo {author} {\bibfnamefont {V.}~\bibnamefont
  {Gorini}}, \bibinfo {author} {\bibfnamefont {A.}~\bibnamefont {Koakowski}},\
  and\ \bibinfo {author} {\bibfnamefont {E.~C.~G.}\ \bibnamefont {Sudarshan}},\
  }\href {https://doi.org/10.1063/1.522979} {\bibfield  {journal} {\bibinfo
  {journal} {Journal of Mathematical Physics}\ }\textbf {\bibinfo {volume}
  {17}},\ \bibinfo {pages} {821} (\bibinfo {year} {1976})}\BibitemShut
  {NoStop}%
\bibitem [{\citenamefont {Lindblad}(1976)}]{Lindblad1976}%
  \BibitemOpen
  \bibfield  {author} {\bibinfo {author} {\bibfnamefont {G.}~\bibnamefont
  {Lindblad}},\ }\href {https://doi.org/10.1007/BF01608499} {\bibfield
  {journal} {\bibinfo  {journal} {Communications in Mathematical Physics}\
  }\textbf {\bibinfo {volume} {48}},\ \bibinfo {pages} {119} (\bibinfo {year}
  {1976})}\BibitemShut {NoStop}%
\bibitem [{\citenamefont {Davies}(1976)}]{Davies1976}%
  \BibitemOpen
  \bibfield  {author} {\bibinfo {author} {\bibfnamefont {E.}~\bibnamefont
  {Davies}},\ }\href {https://books.google.com.br/books?id=I5kuAAAAIAAJ} {
  {\bibinfo {title} {Quantum Theory of Open Systems}}}\ (\bibinfo  {publisher}
  {Academic Pre},\ \bibinfo {year} {1976})\BibitemShut {NoStop}%
\bibitem [{Sch(2007)}]{Schlosshauer2007}%
  \BibitemOpen
  \href {https://doi.org/10.1007/978-3-540-35775-9} {}Frontiers Collection\
  (\bibinfo  {publisher} {Springer Berlin Heidelberg},\ \bibinfo {address}
  {Berlin, Heidelberg},\ \bibinfo {year} {2007})\BibitemShut {NoStop}%
\bibitem [{\citenamefont {Levy}\ and\ \citenamefont
  {Kosloff}(2014)}]{Levy2014}%
  \BibitemOpen
  \bibfield  {author} {\bibinfo {author} {\bibfnamefont {A.}~\bibnamefont
  {Levy}}\ and\ \bibinfo {author} {\bibfnamefont {R.}~\bibnamefont {Kosloff}},\
  }\href {https://doi.org/10.1209/0295-5075/107/20004} {\bibfield  {journal}
  {\bibinfo  {journal} {Europhysics Letters}\ }\textbf {\bibinfo {volume}
  {107}},\ \bibinfo {pages} {20004} (\bibinfo {year} {2014})}\BibitemShut
  {NoStop}%
\bibitem [{\citenamefont {Kosloff}(2013)}]{Kosloff2013}%
  \BibitemOpen
  \bibfield  {author} {\bibinfo {author} {\bibfnamefont {R.}~\bibnamefont
  {Kosloff}},\ }\href {https://doi.org/10.3390/e15062100} {\bibfield  {journal}
  {\bibinfo  {journal} {Entropy}\ }\textbf {\bibinfo {volume} {15}},\ \bibinfo
  {pages} {2100–2128} (\bibinfo {year} {2013})}\BibitemShut {NoStop}%
\bibitem [{\citenamefont {De~Chiara}\  {et~al.}(2018)\citenamefont
  {De~Chiara}, \citenamefont {Landi}, \citenamefont {Hewgill}, \citenamefont
  {Reid}, \citenamefont {Ferraro}, \citenamefont {Roncaglia},\ and\
  \citenamefont {Antezza}}]{DeChiara2018}%
  \BibitemOpen
  \bibfield  {author} {\bibinfo {author} {\bibfnamefont {G.}~\bibnamefont
  {De~Chiara}}, \bibinfo {author} {\bibfnamefont {G.}~\bibnamefont {Landi}},
  \bibinfo {author} {\bibfnamefont {A.}~\bibnamefont {Hewgill}}, \bibinfo
  {author} {\bibfnamefont {B.}~\bibnamefont {Reid}}, \bibinfo {author}
  {\bibfnamefont {A.}~\bibnamefont {Ferraro}}, \bibinfo {author} {\bibfnamefont
  {A.~J.}\ \bibnamefont {Roncaglia}},\ and\ \bibinfo {author} {\bibfnamefont
  {M.}~\bibnamefont {Antezza}},\ }\href
  {https://doi.org/10.1088/1367-2630/aaecee} {\bibfield  {journal} {\bibinfo
  {journal} {New Journal of Physics}\ }\textbf {\bibinfo {volume} {20}},\
  \bibinfo {pages} {113024} (\bibinfo {year} {2018})}\BibitemShut {NoStop}%
\bibitem [{\citenamefont {Landi}\ and\ \citenamefont
  {Paternostro}(2020)}]{Landi2020}%
  \BibitemOpen
  \bibfield  {author} {\bibinfo {author} {\bibfnamefont {G.~T.}\ \bibnamefont
  {Landi}}\ and\ \bibinfo {author} {\bibfnamefont {M.}~\bibnamefont
  {Paternostro}},\ }\href@noop {} {\bibinfo {title} {Irreversible entropy
  production, from quantum to claical}} (\bibinfo {year} {2020}),\ \Eprint
  {https://arxiv.org/abs/2009.07668} {arXiv:2009.07668} \BibitemShut {NoStop}%
\bibitem [{\citenamefont {Colla}\ and\ \citenamefont
  {Breuer}(2022)}]{Colla2022}%
  \BibitemOpen
  \bibfield  {author} {\bibinfo {author} {\bibfnamefont {A.}~\bibnamefont
  {Colla}}\ and\ \bibinfo {author} {\bibfnamefont {H.-P.}\ \bibnamefont
  {Breuer}},\ }\href@noop {} {\bibfield  {journal} {\bibinfo  {journal}
  {Physical Review A}\ }\textbf {\bibinfo {volume} {105}},\ \bibinfo {pages}
  {052216} (\bibinfo {year} {2022})}\BibitemShut {NoStop}%
\bibitem [{\citenamefont {Hayden}\ and\ \citenamefont
  {Sorce}(2022)}]{Hayden2022}%
  \BibitemOpen
  \bibfield  {author} {\bibinfo {author} {\bibfnamefont {P.}~\bibnamefont
  {Hayden}}\ and\ \bibinfo {author} {\bibfnamefont {J.}~\bibnamefont {Sorce}},\
  }\href {https://doi.org/10.1088/1751-8121/ac65c2} {\bibfield  {journal}
  {\bibinfo  {journal} {Journal of Physics A: Mathematical and Theoretical}\
  }\textbf {\bibinfo {volume} {55}},\ \bibinfo {pages} {225302} (\bibinfo
  {year} {2022})}\BibitemShut {NoStop}%
\bibitem [{Note1()}]{Note1}%
  \BibitemOpen
  \bibinfo {note} {Remember that a quotient space is the space where each
  element is one class of equivalence.}\BibitemShut {Stop}%
\bibitem [{\citenamefont {Frigerio}(1977)}]{Frigerio1977}%
  \BibitemOpen
  \bibfield  {author} {\bibinfo {author} {\bibfnamefont {A.}~\bibnamefont
  {Frigerio}},\ }\href {https://doi.org/10.1007/BF00398571} {\bibfield
  {journal} {\bibinfo  {journal} {Letters in Mathematical Physics}\ }\textbf
  {\bibinfo {volume} {2}},\ \bibinfo {pages} {79} (\bibinfo {year}
  {1977})}\BibitemShut {NoStop}%
\bibitem [{\citenamefont {Baumgartner}\ and\ \citenamefont
  {Narnhofer}(2012)}]{Baumgartner2012}%
  \BibitemOpen
  \bibfield  {author} {\bibinfo {author} {\bibfnamefont {B.}~\bibnamefont
  {Baumgartner}}\ and\ \bibinfo {author} {\bibfnamefont {H.}~\bibnamefont
  {Narnhofer}},\ }\href {https://doi.org/10.1142/S0129055X12500018} {\bibfield
  {journal} {\bibinfo  {journal} {Reviews in Mathematical Physics}\ }\textbf
  {\bibinfo {volume} {24}},\ \bibinfo {pages} {1250001} (\bibinfo {year}
  {2012})}\BibitemShut {NoStop}%
\bibitem [{\citenamefont {Albert}\ and\ \citenamefont
  {Jiang}(2014)}]{Albert2014}%
  \BibitemOpen
  \bibfield  {author} {\bibinfo {author} {\bibfnamefont {V.~V.}\ \bibnamefont
  {Albert}}\ and\ \bibinfo {author} {\bibfnamefont {L.}~\bibnamefont {Jiang}},\
  }\href {https://doi.org/10.1103/PhysRevA.89.022118} {\bibfield  {journal}
  {\bibinfo  {journal} {Physical Review A}\ }\textbf {\bibinfo {volume} {89}},\
  \bibinfo {pages} {022118} (\bibinfo {year} {2014})}\BibitemShut {NoStop}%
\bibitem [{\citenamefont {Olmos}\  {et~al.}(2012)\citenamefont {Olmos},
  \citenamefont {Lesanovsky},\ and\ \citenamefont {Garrahan}}]{Olmos2012}%
  \BibitemOpen
  \bibfield  {author} {\bibinfo {author} {\bibfnamefont {B.}~\bibnamefont
  {Olmos}}, \bibinfo {author} {\bibfnamefont {I.}~\bibnamefont {Lesanovsky}},\
  and\ \bibinfo {author} {\bibfnamefont {J.~P.}\ \bibnamefont {Garrahan}},\
  }\href {https://doi.org/10.1103/PhysRevLett.109.020403} {\bibfield  {journal}
  {\bibinfo  {journal} {Physical Review Letters}\ }\textbf {\bibinfo {volume}
  {109}},\ \bibinfo {pages} {020403} (\bibinfo {year} {2012})}\BibitemShut
  {NoStop}%
\bibitem [{\citenamefont {Manzano}\  {et~al.}(2016)\citenamefont
  {Manzano}, \citenamefont {Chuang},\ and\ \citenamefont {Cao}}]{Manzano2016}%
  \BibitemOpen
  \bibfield  {author} {\bibinfo {author} {\bibfnamefont {D.}~\bibnamefont
  {Manzano}}, \bibinfo {author} {\bibfnamefont {C.}~\bibnamefont {Chuang}},\
  and\ \bibinfo {author} {\bibfnamefont {J.}~\bibnamefont {Cao}},\ }\href
  {https://doi.org/10.1088/1367-2630/18/4/043044} {\bibfield  {journal}
  {\bibinfo  {journal} {New Journal of Physics}\ }\textbf {\bibinfo {volume}
  {18}},\ \bibinfo {pages} {043044} (\bibinfo {year} {2016})}\BibitemShut
  {NoStop}%
\bibitem [{\citenamefont {Mortimer}\  {et~al.}(2025)\citenamefont
  {Mortimer}, \citenamefont {Farina}, \citenamefont {Di~Bello}, \citenamefont
  {Jansen}, \citenamefont {Leitherer}, \citenamefont {Mujal},\ and\
  \citenamefont {Acín}}]{Mortimer2025}%
  \BibitemOpen
  \bibfield  {author} {\bibinfo {author} {\bibfnamefont {L.}~\bibnamefont
  {Mortimer}}, \bibinfo {author} {\bibfnamefont {D.}~\bibnamefont {Farina}},
  \bibinfo {author} {\bibfnamefont {G.}~\bibnamefont {Di~Bello}}, \bibinfo
  {author} {\bibfnamefont {D.}~\bibnamefont {Jansen}}, \bibinfo {author}
  {\bibfnamefont {A.}~\bibnamefont {Leitherer}}, \bibinfo {author}
  {\bibfnamefont {P.}~\bibnamefont {Mujal}},\ and\ \bibinfo {author}
  {\bibfnamefont {A.}~\bibnamefont {Acín}},\ }\href
  {https://doi.org/10.1103/hbrt-cn8q} {\bibfield  {journal} {\bibinfo
  {journal} {Physical Review Research}\ }\textbf {\bibinfo {volume} {7}},\
  \bibinfo {pages} {033237} (\bibinfo {year} {2025})}\BibitemShut {NoStop}%
\bibitem [{\citenamefont {Lau}\  {et~al.}(2023)\citenamefont {Lau},
  \citenamefont {Lim}, \citenamefont {Bharti}, \citenamefont {Kwek},\ and\
  \citenamefont {Vinjanampathy}}]{Lau2023}%
  \BibitemOpen
  \bibfield  {author} {\bibinfo {author} {\bibfnamefont {J.~W.~Z.}\
  \bibnamefont {Lau}}, \bibinfo {author} {\bibfnamefont {K.~H.}\ \bibnamefont
  {Lim}}, \bibinfo {author} {\bibfnamefont {K.}~\bibnamefont {Bharti}},
  \bibinfo {author} {\bibfnamefont {L.-C.}\ \bibnamefont {Kwek}},\ and\
  \bibinfo {author} {\bibfnamefont {S.}~\bibnamefont {Vinjanampathy}},\ }\href
  {https://doi.org/10.1103/PhysRevLett.130.240601} {\bibfield  {journal}
  {\bibinfo  {journal} {Phys. Rev. Lett.}\ }\textbf {\bibinfo {volume} {130}},\
  \bibinfo {pages} {240601} (\bibinfo {year} {2023})}\BibitemShut {NoStop}%
\bibitem [{\citenamefont {et. al.}(2024)}]{Mi2024}%
  \BibitemOpen
  \bibfield  {author} {\bibinfo {author} {\bibfnamefont {X.~M.}\ \bibnamefont
  {et. al.}},\ }\href {https://doi.org/10.1126/science.adh9932} {\bibfield
  {journal} {\bibinfo  {journal} {Science}\ }\textbf {\bibinfo {volume}
  {383}},\ \bibinfo {pages} {1332} (\bibinfo {year} {2024})}\BibitemShut
  {NoStop}%
\bibitem [{\citenamefont {Guo}\  {et~al.}(2025)\citenamefont {Guo},
  \citenamefont {Hart}, \citenamefont {Chen}, \citenamefont {Friedman},\ and\
  \citenamefont {Lucas}}]{GuoHartChen2025}%
  \BibitemOpen
  \bibfield  {author} {\bibinfo {author} {\bibfnamefont {J.}~\bibnamefont
  {Guo}}, \bibinfo {author} {\bibfnamefont {O.}~\bibnamefont {Hart}}, \bibinfo
  {author} {\bibfnamefont {C.-F.}\ \bibnamefont {Chen}}, \bibinfo {author}
  {\bibfnamefont {A.~J.}\ \bibnamefont {Friedman}},\ and\ \bibinfo {author}
  {\bibfnamefont {A.}~\bibnamefont {Lucas}},\ }\href
  {https://doi.org/10.22331/q-2025-01-28-1612} {\bibfield  {journal} {\bibinfo
  {journal} {Quantum}\ }\textbf {\bibinfo {volume} {9}},\ \bibinfo {pages}
  {1612} (\bibinfo {year} {2025})}\BibitemShut {NoStop}%
\bibitem [{\citenamefont {da~Silva~Souza}\ and\ \citenamefont
  {Iemini}(2025)}]{SouzaIemini2024}%
  \BibitemOpen
  \bibfield  {author} {\bibinfo {author} {\bibfnamefont {L.}~\bibnamefont
  {da~Silva~Souza}}\ and\ \bibinfo {author} {\bibfnamefont {F.}~\bibnamefont
  {Iemini}},\ }\href {https://arxiv.org/abs/2408.05302} {\bibinfo {title}
  {Lindbladian reverse engineering for general non-equilibrium steady states: A
  scalable null-space approach}} (\bibinfo {year} {2025}),\ \Eprint
  {https://arxiv.org/abs/2408.05302} {arXiv:2408.05302} \BibitemShut {NoStop}%
\bibitem [{\citenamefont {Zhang}\ and\ \citenamefont
  {Fan}(2015)}]{ZhangFan2015}%
  \BibitemOpen
  \bibfield  {author} {\bibinfo {author} {\bibfnamefont {Y.-R.}\ \bibnamefont
  {Zhang}}\ and\ \bibinfo {author} {\bibfnamefont {H.}~\bibnamefont {Fan}},\
  }\href {https://doi.org/10.1038/srep11509} {\bibfield  {journal} {\bibinfo
  {journal} {Scientific Reports}\ }\textbf {\bibinfo {volume} {5}},\ \bibinfo
  {pages} {11509} (\bibinfo {year} {2015})}\BibitemShut {NoStop}%
\bibitem [{\citenamefont {Toscano}\  {et~al.}(2005)\citenamefont
  {Toscano}, \citenamefont {de~Matos~Filho},\ and\ \citenamefont
  {Davidovich}}]{Toscano2005}%
  \BibitemOpen
  \bibfield  {author} {\bibinfo {author} {\bibfnamefont {F.}~\bibnamefont
  {Toscano}}, \bibinfo {author} {\bibfnamefont {R.~L.}\ \bibnamefont
  {de~Matos~Filho}},\ and\ \bibinfo {author} {\bibfnamefont {L.}~\bibnamefont
  {Davidovich}},\ }\href {https://doi.org/10.1103/PhysRevA.71.010101}
  {\bibfield  {journal} {\bibinfo  {journal} {Physical Review A}\ }\textbf
  {\bibinfo {volume} {71}},\ \bibinfo {pages} {010101} (\bibinfo {year}
  {2005})}\BibitemShut {NoStop}%
\bibitem [{\citenamefont {Wisniacki}\ and\ \citenamefont
  {Toscano}(2009)}]{Toscano2009}%
  \BibitemOpen
  \bibfield  {author} {\bibinfo {author} {\bibfnamefont {D.~A.}\ \bibnamefont
  {Wisniacki}}\ and\ \bibinfo {author} {\bibfnamefont {F.}~\bibnamefont
  {Toscano}},\ }\href {https://doi.org/10.1103/PhysRevE.79.025203} {\bibfield
  {journal} {\bibinfo  {journal} {Physical Review E}\ }\textbf {\bibinfo
  {volume} {79}},\ \bibinfo {pages} {025203} (\bibinfo {year}
  {2009})}\BibitemShut {NoStop}%
\bibitem [{\citenamefont {Hernández}\  {et~al.}(2025)\citenamefont
  {Hernández}, \citenamefont {Ranard},\ and\ \citenamefont
  {Riedel}}]{Ranard2025}%
  \BibitemOpen
  \bibfield  {author} {\bibinfo {author} {\bibfnamefont {F.}~\bibnamefont
  {Hernández}}, \bibinfo {author} {\bibfnamefont {D.}~\bibnamefont {Ranard}},\
  and\ \bibinfo {author} {\bibfnamefont {C.~J.}\ \bibnamefont {Riedel}},\
  }\href {https://doi.org/10.1007/s00220-024-05146-9} {\bibfield  {journal}
  {\bibinfo  {journal} {Communications in Mathematical Physics}\ }\textbf
  {\bibinfo {volume} {406}},\ \bibinfo {pages} {4} (\bibinfo {year}
  {2025})}\BibitemShut {NoStop}%
\bibitem [{Note2()}]{Note2}%
  \BibitemOpen
  \bibinfo {note} {Where the adjoint superoperator is defined with respect to
  the Hilbert-Schmidt scalar product \protect \textit {i.e., }$\langle \protect
  \bar {{\protect \cal L}}(\protect \hat A),\protect \hat \rho \rangle
  _{\protect \rm HS}=\langle \protect \hat A,{\protect \cal L}(\protect \hat
  \rho )\rangle _{\protect \rm HS}$ with $\langle \protect \hat B,\protect \hat
  A\rangle _{\protect \rm HS}=\Tr (\protect \hat B^\dagger \protect \hat
  A)$.}\BibitemShut {Stop}%
\bibitem [{\citenamefont {Macfarlane}\  {et~al.}(1968)\citenamefont
  {Macfarlane}, \citenamefont {Sudbery},\ and\ \citenamefont
  {Weisz}}]{Macfarlane1968}%
  \BibitemOpen
  \bibfield  {author} {\bibinfo {author} {\bibfnamefont {A.~J.}\ \bibnamefont
  {Macfarlane}}, \bibinfo {author} {\bibfnamefont {A.}~\bibnamefont
  {Sudbery}},\ and\ \bibinfo {author} {\bibfnamefont {P.~H.}\ \bibnamefont
  {Weisz}},\ }\href {https://doi.org/10.1007/BF01654302} {\bibfield  {journal}
  {\bibinfo  {journal} {Communications in Mathematical Physics}\ }\textbf
  {\bibinfo {volume} {11}},\ \bibinfo {pages} {77–90} (\bibinfo {year}
  {1968})}\BibitemShut {NoStop}%
\bibitem [{\citenamefont {Haber}(2021)}]{Haber2021}%
  \BibitemOpen
  \bibfield  {author} {\bibinfo {author} {\bibfnamefont {H.}~\bibnamefont
  {Haber}},\ }\href {https://doi.org/10.21468/SciPostPhysLectNotes.21}
  {\bibfield  {journal} {\bibinfo  {journal} {SciPost Physics Lecture Notes}\
  ,\ \bibinfo {pages} {21}} (\bibinfo {year} {2021})}\BibitemShut {NoStop}%
\bibitem [{\citenamefont {vom Ende}(2024)}]{vomEnde2024}%
  \BibitemOpen
  \bibfield  {author} {\bibinfo {author} {\bibfnamefont {F.}~\bibnamefont {vom
  Ende}},\ }\href {https://doi.org/10.1142/S1230161224500070} {\bibfield
  {journal} {\bibinfo  {journal} {Open Systems \& Information Dynamics}\
  }\textbf {\bibinfo {volume} {31}},\ \bibinfo {pages} {2450007} (\bibinfo
  {year} {2024})}\BibitemShut {NoStop}%
\bibitem [{\citenamefont {Spohn}\ and\ \citenamefont
  {Lebowitz}(1978)}]{SpohnLebowitz1978}%
  \BibitemOpen
  \bibfield  {author} {\bibinfo {author} {\bibfnamefont {H.}~\bibnamefont
  {Spohn}}\ and\ \bibinfo {author} {\bibfnamefont {J.~L.}\ \bibnamefont
  {Lebowitz}},\ }\bibinfo {title} {Irreversible thermodynamics for quantum
  systems weakly coupled to thermal reservoirs},\ in\ \href@noop {} {
  {\bibinfo {booktitle} {Advances in Chemical Physics}}}\ (\bibinfo
  {publisher} {John Wiley \& Sons, Ltd},\ \bibinfo {year} {1978})\ pp.\
  \bibinfo {pages} {109--142}\BibitemShut {NoStop}%
\bibitem [{\citenamefont {Alicki}(1976)}]{Alicki1976}%
  \BibitemOpen
  \bibfield  {author} {\bibinfo {author} {\bibfnamefont {R.}~\bibnamefont
  {Alicki}},\ }\href {https://doi.org/10.1016/0034-4877(76)90046-X} {\bibfield
  {journal} {\bibinfo  {journal} {Reports on Mathematical Physics}\ }\textbf
  {\bibinfo {volume} {10}},\ \bibinfo {pages} {249–258} (\bibinfo {year}
  {1976})}\BibitemShut {NoStop}%
\bibitem [{Note3()}]{Note3}%
  \BibitemOpen
  \bibinfo {note} {Details will be published elsewhere.}\BibitemShut {Stop}%
\bibitem [{\citenamefont {Hall}\  {et~al.}(2014)\citenamefont {Hall},
  \citenamefont {Creer}, \citenamefont {Li},\ and\ \citenamefont
  {Anderon}}]{Hall2014}%
  \BibitemOpen
  \bibfield  {author} {\bibinfo {author} {\bibfnamefont {M.~J.~W.}\
  \bibnamefont {Hall}}, \bibinfo {author} {\bibfnamefont {J.~D.}\ \bibnamefont
  {Creer}}, \bibinfo {author} {\bibfnamefont {L.}~\bibnamefont {Li}},\ and\
  \bibinfo {author} {\bibfnamefont {E.}~\bibnamefont {Anderon}},\ }\href
  {https://doi.org/10.1103/PhysRevA.89.042120} {\bibfield  {journal} {\bibinfo
  {journal} {Physical Review A}\ }\textbf {\bibinfo {volume} {89}},\ \bibinfo
  {pages} {042120} (\bibinfo {year} {2014})}\BibitemShut {NoStop}%
\end{thebibliography}


%

\clearpage
\appendix
\onecolumngrid
\section{Derivation of Eqs.(13) and (14) of the main manuscript.}
\label{App-deriv1}
\indent 

Here we used the expansion of the Lindblad operators $\hL_k=\sum_\nu \alk_{\nu}\hT_\nu$ in all the derivations.
In order to derive Eqs.(13) of the main manuscript note that 
\bse
\begin{align}
\hA_k&=\frac{1}{2}(\hL_k+\hL_k^\dagger)=\frac{1}{2}\sum_{\nu=0}^{N^2-1} (\alpha^{(k)}_{\nu} + \alpha^{*(k)}_\nu )\hT_\nu=\sum_\nu \Re(\alk_\nu)\hT_\nu,\\
\hB_k&=\frac{1}{2\imath}(\hL_k-\hL_k^\dagger)=\frac{1}{2\imath}\sum_{\nu=0}^{N^2-1} (\alpha^{(k)}_{\nu}-\alpha^{*(k)}_\nu)\hT^\dagger_\nu=\sum_\nu\Im(\alk_\nu)\hT_\nu.
\end{align}
\ese
 So, we can write: 
 \bse
\begin{align}
 [\hA_k  , [\hA_k  , \cdot]]=
\sum_\nu\sum_\mu \Re(\alk_\nu) \Re(\alk_\mu)
[\hT_\nu,[\hT_\mu,\cdot]]=\sum_i\sum_j \Re(\alk_i) \Re(\alk_j)[\hT_i,[\hT_j,\cdot]],\\
 [\hB_k  , [\hB_k  , \cdot]]= \sum_\nu\sum_\mu \Im(\alk_\nu) \Im(\alk_\mu)
[\hT_\nu,[\hT_\mu,\cdot]]=\sum_i\sum_j \Im(\alk_i) \Im(\alk_j)
[\hT_i,[\hT_j,\cdot]].
\end{align}
\ese
Therefore, using $\Re(zw^*)=\Re(z)\Re(w)+\Im(z)\Im(w)$ we recover Eq.(13a), \ie
\begin{gather}
\calL_1(\hO)=-\frac{1}{2\hbar}\sum_{k=1}^K\left([\hA_k  , [\hA_k  , \hO]]+[\hB_k  , [\hB_k  , \hO]]\right)=-\frac{1}{2\hbar}\sum_i\sum_j \Re
\left(\sum_{k=1}^K\alk_\nu\alkb_\mu\right)
[\hT_i,[\hT_j^\dagger,\hO]]=\nonumber\\
=-\frac{1}{2\hbar^2}\sum_i\sum_j \mD_{ij}
[\hT_i,[\hT_j^\dagger,\hO]].
\end{gather}
\par
In order to derive Eq.(13b) note that we can write:
\begin{align}
\calL_2(\cdot)&=\sum_{k=1}^K\frac{1}{2}\{[\hL_k,\hL_k^\dagger],\cdot\}=\frac{1}{2}\left(\sum_{k=1}^K[\hL_k,\hL_k^\dagger]\cdot+\cdot\sum_{k=1}^K[\hL_k,\hL_k^\dagger]\right)\nonumber\\
&=\frac{1}{2}\sum_{i=1}^{N^2-1}\sum_{j=1}^{N^2-1}\mGam_{ij}\,\left([\hT_i,\hT_j]\cdot+\cdot[\hT_i,\hT_j]\right)\nonumber\\
&=
\frac{1}{2}\sum_{i=1}^{N^2-1}\sum_{j=1}^{N^2-1}\mGam_{ij}\hT_i\hT_j\cdot
-\frac{1}{2}\sum_{i=1}^{N^2-1}\sum_{j=1}^{N^2-1}\mGam_{ij}\hT_j\hT_i\cdot\nonumber\\
&+\frac{1}{2}\sum_{i=1}^{N^2-1}\sum_{j=1}^{N^2-1}\mGam_{ij}\cdot\hT_i\hT_j
-\frac{1}{2}\sum_{i=1}^{N^2-1}\sum_{j=1}^{N^2-1}\mGam_{ij}\cdot \hT_j\hT_i.
\end{align}
Now, using $\mGam^\dagger_{ij}=\mGam_{ij}=\mGam^*_{ji}$ we can write:
\begin{align}
\frac{1}{2}\sum_i\sum_j\mGam_{ij}\hT_j\hT_i\cdot&=
\frac{1}{2}\sum_i\sum_j\mGam_{ij}^*\hT_j\hT_i\cdot=
\frac{1}{2}\sum_i\sum_j\mGam_{ij}^*\hT_i\hT_j\cdot\;,\\
\frac{1}{2}\sum_i\sum_j\mGam_{ij}\cdot\hT_j\hT_i&=
\frac{1}{2}\sum_i\sum_j\mGam_{ij}^*\cdot\hT_j\hT_i=
\frac{1}{2}\sum_i\sum_j\mGam_{ij}^*\cdot\hT_i\hT_j
\end{align}
where in the last equality we change the indexes $i\leftrightarrow j$.
So, we have that 
\begin{align}
\calL_2(\cdot)&=\frac{1}{2\hbar}\sum_{k=1}^K\frac{1}{2}\{[\hL_k,\hL_k^\dagger],\cdot\}=\frac{\imath}{2\hbar}
\sum_i\sum_j
\frac{(\mGam_{ij}-\mGam_{ij}^*)}{2\imath}\{\hT_i\hT_j,\cdot\}\nonumber\\
&=\frac{\imath}{2\hbar}\sum_i\sum_j\Im(\mGam_{ij})\{\hT_i\hT_j,\cdot\}=
\frac{1}{\hbar}\sum_i\sum_j\imath\mC_{ij}\frac{1}{2}\{\hT_i\hT_j,\cdot\}\;.
\end{align}
\par
In order to derive Eq.(13c), we first write: 
\begin{align}
\nonumber
\sum_{k=1}^K[\hL_k,\hL_k^\dagger]&=\sum_\nu\sum_\mu\sum_{k=1}^K\alk_\nu\alkb_\mu\,[\hT_\nu\hT_\mu]=
\sum_\nu\sum_\mu\mGamt_{\nu\mu}\,[\hT_\nu,\hT_\mu]=\sum_i\sum_j\mGam_{ij}\,[\hT_i,\hT_j]=\\
\nonumber
&=2\imath\sum_l\left(\sum_i\sum_j\,\mGam_{ij}\,f_{ijl}\right)\hT_l=2\sum_l
\left(\sum_i\sum_j\,\imath\mC_{ij}\,\imath f_{ijl}\right)\,\hT_l=\\
&=2\sum_l
\left(\sum_i\sum_j\,\imath\mC_{ij}\,\mf^l_{ji}\right)\,\hT_l=2\sum_l\tr(\imath\mC\mf^l)\hT_l,
\label{eq:LkLkcomu}
\end{align}
where we used $ [\hT_i,\hT_j]=\imath2\sum_lf_{ijl}\hT_l$, and that $\mGam_{ij}=\frac{\mD_{ij}}{\hbar}+\imath \mC_{ij}$, so
$\sum_i\sum_j\mD_{ij}f_{ijl}=0$. Then, replacing~\eqref{eq:LkLkcomu} into
$\calL_2(\cdot)  = \frac{1}{2\hbar}  \sum_{k=1}^K \frac12 \{ [\hL_k ,\hL_k^\dag]
, \cdot \}$ we obtain Eq.(13c).
\par
\par
In order to derive Eq.(13d), using $\hL_k=\sum_\nu \alk_{\nu}\hT_\nu$
into the definition of $\calL_3$, we can write: 
\begin{align}
\nonumber
\calL_3(\cdot)&=\frac{1}{2\hbar}  \sum_{k=1}^K  (\hL_k  \cdot  \hL^\dag_k-\hL_k^\dag \cdot \hL_k)=\frac{1}{\hbar}\sum_\nu\sum_\mu\frac{1}{2}\left(\sum_{k=1}^K\alk_\nu\alkb_\mu-\sum_{k=1}^K\alkb_\nu\alk_\mu\right) \hT_\nu\cdot\hT_\mu =\\
&=\frac{1}{\hbar}\sum_\nu\sum_\mu \imath(\mCt)_{\nu\mu}\hT_\nu\cdot\hT_\mu =
-\frac{1}{\hbar}\sum_\mu
\imath(\mCt)_{0\mu}\sqrt{\frac{2}{N}}(\hT_\mu\cdot\hid-\hid\cdot\hT_\mu)+\frac{1}{\hbar}
\sum_i\sum_j
\imath(\mC)_{ij}\hT_i\cdot\hT_j=\nonumber\\
&=
-\frac{\imath}{\hbar}\left[\sqrt{\frac{2}{N}}\sum_i
(\mCt)_{0i}\hT_i,\cdot\right]+\frac{1}{\hbar}
\sum_i\sum_j
\imath(\mC)_{ij}\hT_i\cdot\hT_j\nonumber\\
&=-\frac{\imath}{\hbar}\left[\hHb,\cdot\right]+\calL^\prime_3(\cdot),
\end{align}
where $(\vecC)_i=(\mCt)_{0i}$ and we define the Hamiltonian: 
\begin{align}
\hHb =-\frac{1}{2\imath\,N }\sum_{k=1}^K   \left( \Tr(\hL_k)^*  \hL_k  -    \Tr(\hL_k)   \hL_k^\dag\right)= 
-\frac{1}{2\imath }\sqrt{\frac{2}{N}}\sum_\nu\sum_{k=1}^K   (  \alpha_0^{(k)*}   \alpha_\nu^{(k)}  -   \alpha_0^{(k)}  \alpha_\nu^{(k)*})\hT_\nu=\nonumber\\
=\sqrt{\frac{2}{N}}\sum_i\Im\left(\sum_{k=1}^K  \alpha_0^{(k)}   \alpha_i^{(k)*} \right)\hT_i
=\sqrt{\frac{2}{N}}\sum_i(\mCt)_{0i}\hT_i=\sqrt{\frac{2}{N}}\sum_i(\vecC)_{i}\hT_i,
\end{align}
with $\Tr(\hL_k)=\alpha_0^{(k)}\sqrt{\frac{2}{N}}N$. Therefore we have obtained the results in Eq.(14)
and~(13d), \ie $\calL_3^\prime(\cdot)=
\frac{1}{\hbar}
\sum_i\sum_j
\imath\,\mC_{ij}\hT_i\cdot\hT_j$.

\end{document}